\documentclass[final,5p,times,twocolumn]{elsarticle}

\usepackage{amssymb}
\usepackage{amsthm}

\usepackage{hyperref}
\usepackage{siunitx} 
\usepackage{subfigure}
\usepackage{algorithm}
\usepackage{algpseudocode}

\usepackage{ifthen}
\usepackage{calc}
\usepackage{fancybox}
\theoremstyle{remark}
\newtheorem*{Todo}{TODO}
	{\vspace*{0.5ex}\ifthenelse{\equal{#1}{}}
		{\noindent\begin{Sbox}\begin{minipage}{\columnwidth-1ex}\begin{Todo}}%
		{\noindent\begin{Sbox}\begin{minipage}{\columnwidth-1ex}\begin{Todo}[#1]}%
	}%
	{\end{Todo}\end{minipage}\end{Sbox}\fbox{\TheSbox}\vspace*{0.5ex}}

\journal{arxiv.org}

\begin{document}

\begin{frontmatter}



\title{Automatic trajectory recognition in Active Target Time Projection Chambers data by means of hierarchical clustering}


\author[a,1]{Christoph Dalitz}
\ead{christoph.dalitz@hsnr.de}
\author[b,1]{Yassid Ayyad}
\ead{ayyad@lbl.gov}
\author[a]{Jens Wilberg}
\author[a]{Lukas Aymans}
\author[c]{Daniel Bazin}
\author[c]{and Wolfgang Mittig}

\address[a]{Niederrhein University of Applied Sciences, Institute for Pattern Recognition\\Reinarzstr. 49, 47805 Krefeld, Germany}
\address[b]{Lawrence Berkeley National Laboratory, Berkeley, CA 94720, USA}
\address[c]{National Superconducting Cyclotron Laboratory,\\640 S. Shaw Lane, East Lansing, MI 48824, USA}


\begin{abstract}
The automatic reconstruction of three-dimensional particle tracks from Active Target Time Projection Chambers data can be a challenging task, especially in the presence of noise. In this article, we propose a non-parametric algorithm that is based on the idea of clustering point triplets instead of the original points. We define an appropriate distance measure on point triplets and then apply a single-link hierarchical clustering on the triplets. Compared to parametric approaches like RANSAC or the Hough transform, the new algorithm has the advantage of potentially finding trajectories even of shapes that are not known beforehand. This feature is particularly important in low-energy nuclear physics experiments with Active Targets operating inside a magnetic field. The algorithm has been validated using data from experiments performed with the Active Target Time Projection Chamber developed at the National Superconducting Cyclotron Laboratory (NSCL).The results demonstrate the capability of the algorithm to identify and isolate particle tracks that describe non-analytical trajectories. For curved tracks, the vertex detection recall was 86\% and the precision 94\%. For straight tracks, the vertex detection recall was 96\% and the precision 98\%. In the case of a test set containing only straight linear tracks, the algorithm performed better than an iterative Hough transform.
\end{abstract}


\end{frontmatter}


\section{Introduction}
\label{sec:intro}

One of the present aims of modern low-energy nuclear physics is to provide a more complete understanding about the behavior of subatomic matter under large isospin (i.e. large neutron-proton imbalance)~\cite{Thoennessen2011}. Near the landscape drip lines, atomic nuclei exhibit astonishing properties such as disappearance of magic numbers, exotic collective modes (i.e. development of neutron/proton halos and giant resonances) or a rearrangement of the matter in cluster and molecular structures. In addition, the vast scenario that composes the nucleosynthesis of heavy elements through different astrophysical processes, primarily involves very exotic nuclei. Aware of the impact that this research will pose, the nuclear physics community is striving to upgrade and construct state-of-the-art facilities worldwide in order to achieve the required exotic beam intensities that would allow for a comprehensive and efficient study of the unexplored region of the landscape~\cite{Motobayashi2014}.

The production of exotic beams also involves the development of associated instrumentation capable of providing the observables of interest in a reasonable amount of time for the expected rates. Active Target Time Projection Chambers (henceforth AT) detectors are designed to provide high luminosity (rate of interactions) while preserving the resolution needed to obtain the variables from which relevant information about the nuclei is inferred~\cite{BeceiroNovo2015124}. ATs consist of a relatively large gaseous medium that acts as target and detector at the same time. They provide a target thickness that could be up to two orders of magnitude larger than conventional solid targets and also the capability of recording three-dimensional particle tracks with high geometrical efficiency covering a solid angle of 4$\pi$. In this sense, ATs detectors are able to push the rate limits at which low-energy nuclear physics experiments can be performed, with intensities as low as 100 pps (particle per second). Moreover, placing the detector inside a magnetic field enables high-resolution spectrometer operation~\cite{BRADT201765,WUOSMAA20071290}.

Contrary to High Energy Physics (HEP) experiments, nuclear reactions at low energy (in the range of few hundreds of keV up to several hundred MeV) produce charged particles in a broad range of atomic number ($Z$) and mass ($A$), and also with very different kinetic energy. Since the reaction of interest occurs inside the AT medium, the detector has to cope with a large energy dynamic range which is directly reflected on the distribution of patterns and shapes of tracks that the particles describe in the gas under a magnetic field. While particles with relatively high energy will describe helical trajectories (projections can be considered circular), the ones with low energy will describe non-linear trajectories as they are progressively stopped in the gas. In addition, particle tracks are connected through the vertex of the reaction from which the energy of the beam at the interaction point can be inferred.

Generally speaking, ATs are versatile detectors that can be deployed in different experiments covering a wide variety of reactions such as resonant scattering or nucleon transfer, that are normally performed using beams with energies from few \si{\mega\electronvolt} up to several tens of \si{A.\mega\electronvolt}~\cite{MITTIG2015494,BeceiroNovo2015124,MITTIG2001495}. The particle multiplicity in these reactions is relatively low, from two up to several particles, depending on the physics case of interest. In order to reconstruct the kinematics of the reaction completely, the information from the beam and from each reaction product has to be extracted. That includes the angles with respect to the beam direction, and their range and/or their curvature, depending if a magnetic field was used. Other types of reactions for which ATs are one of the most promising approaches are elastic/inelastic scattering and charge exchange with fast beams (more than \SI{100}{A.\mega\electronvolt}). In this case, the relevant kinematic information is obtained from particles that are emitted at very forward angles in center-of-mass which essentially implies very low kinetic energy (or short range).

Although ATs are the answer to many experimental demands in low-energy nuclear physics experiments, the analysis of the data is a complicated and demanding task that needs to be performed in different steps. Before performing any comprehensive kinematic analysis, one needs to break down every image recorded by the AT in order to extract different features. The purpose is twofold: identify events where a reaction happened and isolate every particle track. Due to the large amount of collected data (data streaming during an experiment can be of the order of hundreds MB/sec), and efficient and fast discrimination of useful events is mandatory. The requirements are quite restrictive: The algorithm needs to isolate every track, even very short ones, while preserving the information about the reaction vertex.

A common approach for detecting shapes in noisy point clouds consists in a parametric description of the shapes and to search for parameter values representing shapes with many points. This requires the shape of the tracks to be known a priori, and it leads to a voting scheme for parameter values, which can be either done exhaustively for all points (Hough transform \cite{jeltsch16}) or by random sampling (RANSAC \cite{fischler87}). In the absence of a magnetic field, the tracks are straight lines, which can easily be described parametrically, and both RANSAC and the Hough transform have been applied to this special case \cite{dalitz17,ayyad18}. This is not generalizable to the case of unknown shapes like  they occur in the presence of a magnetic field.

We therefore present a new, non-parametric approach in the present article. It is based on first building point triplets, which are then partitioned into clusters, such that each cluster represents a track. The idea of grouping triplets was recently suggested by Lezama et al.~for detecting continuations in 2D dot patterns \cite{lezama17}. Their algorithm looks for symmetric triplets only, and the grouping is based on overlapping points. Although that algorithm could be generalized to 3D in a straightforward way, it is applicable only in the special case of locally equidistant points without random perpendicular spread around the actual curve. As these assumptions do not hold in the case of AT data, we change both steps of the algorithm: firstly, we do not look for symmetric triplets, but for triplets with approximately collinear branches, and, secondly, we group the triplets with a single link hierarchical clustering utilizing an appropriately defined triplet distance metric.

The algorithm has been developed within the framework of the AT detector developed at the NSCL (the so-called AT-TPC), which operates inside a solenoid that provides a magnetic field up to 2~T~\cite{BRADT201765}. The source code of this framework is available from {\em github}\footnote{\url{https://github.com/ATTPC/ATTPCROOTv2}}. Considering the conditions in which low-energy physics experiments are performed, the particles of interest describe trajectories that cannot be described by an analytical expression in closed form. This algorithm has been tailored to extract such curved tracks described by every reaction partner recorded on a event-by-event basis. Once the tracks are isolated, a Monte Carlo fit is applied to extract the variables of interest with better precision~\cite{BRADT201765}. The fit makes use of the parameters of each curve, namely curvature and angle with respect to the beam direction, to numerically generate a collection of simulated tracks that are compared to the experimental one using an objective function. Therefore, the algorithm must provide the initial parameters with enough precision to ensure the proper convergence of the fit. With a broad experimental program\footnote{A list of approved experiments at the NSCL can be found here: \url{https://enterprise.nscl.msu.edu/completedExperiments/} (e18027, e18019, e18008, e17504, e17025, e15250, e15238)}, the development of a generic and fast algorithm that addresses this unprecedented situation is mandatory for low-energy physics experiments using ATs in magnetic fields.

The present article is organized as follows: in section \ref{sec:attpc}, we give a brief overview of the AT measurement technique, and in section \ref{sec:data} we describe the three experimental data sets used throughout this article. Section \ref{sec:algorithm} describes the algorithm in detail, section \ref{sec:evaluation} presents its performance on our test data sets, and section \ref{sec:conclusions} makes recommendations for practical deployments of the algorithm.

\section{Active target time projection chambers}
\label{sec:attpc}

\begin{figure}
\center
\includegraphics[width=1.0\columnwidth]{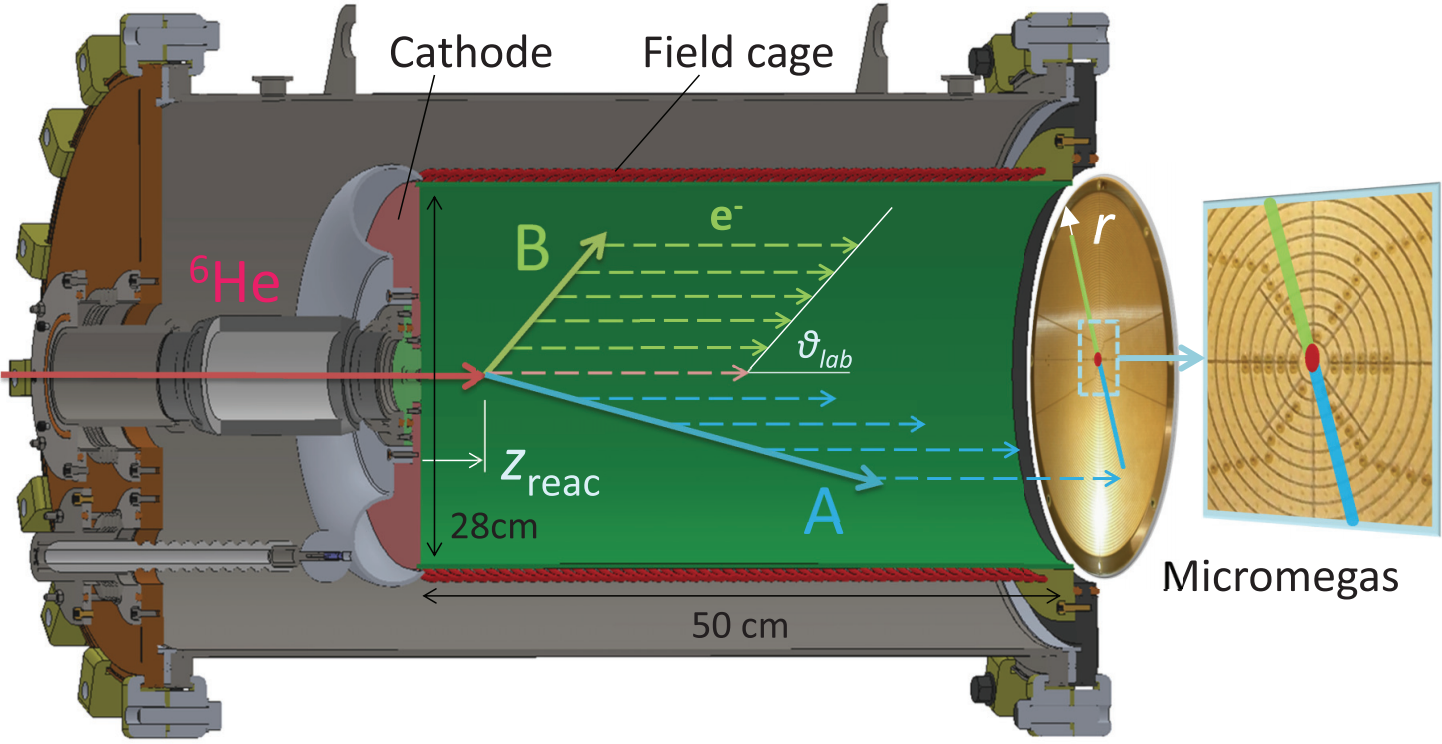}
\caption{\label{fig:pATTPC}Schematic view of the prototype AT-TPC (pAT-TPC). The solid  and dashed arrows represent particles and the ionization electrons, respectively. $Z_{\mbox{\scriptsize\em reac}}$ denotes the vertex of the reaction. More details can be found in Ref.~\cite{PhysRevC.87.054301}.}
\end{figure}

The working principle of ATs is the same as Time Projection Chambers~\cite{Nygren_1974} with the particularity that the tracking medium acts as the target. This limits the type of gas used for tracking since typical experiments in nuclear physics are performed with proton, deuteron, $^{3}$He or $^{4}$He targets, among others. The detector consists of gas volume in which a strong uniform electric field (typical values are \SI{100}{V/cm}) is applied along one of the volume coordinates. The uniform electric field is created by a set of electrodes defining a drift volume (field cage) with a geometry defined by the requirements of the detector. Ionization electrons released by particles that cross the volume, drift towards the anode electrode where a segmented readout plane is deployed. Therefore, each ionization electron cloud is projected onto a pad of the plane from where the $x$ and $y$ coordinates can be determined. The third $z$ coordinate is deduced from the drift time of the electrons assuming a constant drift velocity that depends on the gas and on the voltage. The pads also collect the total charge which is proportional to the local energy loss. Fig.~\ref{fig:pATTPC} illustrates the main working principle of ATs using the prototype AT-TPC (pAT-TPC) as an example~\cite{PhysRevC.87.054301,Suzuki12}. The detector has a cylindrical configuration with the electric field applied between the cathode and the pad plane (right end). The beam enters the chamber through the cathode end and it is slowed down in the gas. When a reaction takes place with one of the target nuclei ($Z_{\mbox{\scriptsize\em reac}}$ refers to the reaction vertex), reaction products are emitted with a characteristic angle and energy distribution. Every particle track is projected into the pad plane as explained before: The beam and the products are projected into a small spot and a straight line, respectively, as can be seen in the figure. The energy of the reaction products can be inferred from the track length and from their characteristic energy loss pattern.

Depending on the reaction of interest and also on the energy of the beam, particles with relatively large energy can escape from the detector. In this case, the identification of the particle and the determination of its energy is not possible anymore because part of the track is lost. As consequence, the total efficiency of the detector is reduced. In order to extend the energy dynamic range of the detector and preserve its excellent efficiency, one of the solutions is to place the AT inside a solenoid magnet. This enables the measurement of the magnetic rigidity of the particle by determining the radius of curvature of the track. The energy of the particle (and also its nature) can be determined with a very simple relationship:

\begin{equation}
B\rho = \frac{p}{q}
\end{equation}

where $B$ is the magnetic field, $\rho$ is the radius of curvature, $p$ is the momentum of the particle (from where the energy is inferred) and $q$ is its charge. The AT-TPC, which is an improved version of the pAT-TPC featuring larger volume and higher granularity, is presently the only AT operated under a magnetic field. The detector consist of a cylindrical gas volume of length \SI{1}{m} and \SI{29.2}{cm} of radius with a pad plane segmented in triangular pads of \SI{0.5}{cm} (inner region) and \SI{1.0}{cm} (outer region) of height. The detector is placed inside a large-bore solenoid magnet capable of producing a uniform magnetic field up to \SI{2}{T}. Further details about the detector can be found in Ref.~\cite{BRADT201765}. The detector was successfully commissioned using the $^{4}$He+$^{4}$He (detector filled with $^{4}$He gas~\cite{ayyad18}) and $^{46}$Ar+proton (detector filled with C$_4$H$_{10}$ gas)~\cite{Bradt2018} reactions in inverse kinematics. These two reactions, which represent typical experiments in low-energy nuclear physics, provide very different benchmark points: In the former, the angle between both identical $^{4}$He particles, which amounts to \SI{90}{\degree} exactly, can be used to infer the resolution of the detector and of the tracking method. The latter reaction provides a rather different scenario: The mass of the $^{46}$Ar is approximately 46 times larger than that of the proton, and therefore the tracks of the particles and their energy loss profiles exhibit a very different topology. In either case, the trajectories cannot be described by analytical expressions, which emphasizes the need of a novel tracking algorithm better suited to this type of experiments using the emerging AT technology in worldwide nuclear physics facilities.

\section{Data description}
\label{sec:data}

The image of each track is recorded by measuring the drifting electrons that arrive to the highly-segmented pad plane (10,240 pads in total). The charge collected in each pad generates a pulse whose amplitude is proportional to the charge. The pulses are digitized in parallel using a dedicated and generic data-acquisition system (General Electronics for TPCs~\cite{Pollacco201881}) consisting of 40 ASIC boards featuring 256 12-bit ADC (4096 channels). The pulse is sampled with a maximum frequency of \SI{100}{MHz} and 512 sampling time bins. The centroid of each pulse is used to extract the mean time ($t_{\mbox{\scriptsize\em drift}}$) that is used to infer the $z$ coordinate using a simple relationship:

\begin{equation}
z = v_{\mbox{\scriptsize\em drift}} \cdot t_{\mbox{\scriptsize\em drift}}
\end{equation}

where $v_{\mbox{\scriptsize\em drift}}$ is the drift velocity of the ionization electrons. $v_{\mbox{\scriptsize\em drift}}$ is about 2.0 and \SI{5.20}{cm/\micro s} for $^{4}$He and C$_{4}$H$_{10}$ gases, respectively. Finally, the $x$ and $y$ coordinates are inferred from the centroid of the triangular pads with 0.5 and \SI{1.0}{cm} of height, deployed in the inner and outer region of the pad plane, respectively.

\begin{figure}[t!]
\center
\subfigure[Sketch of the AT-TPC]{\includegraphics[width=0.80\columnwidth]{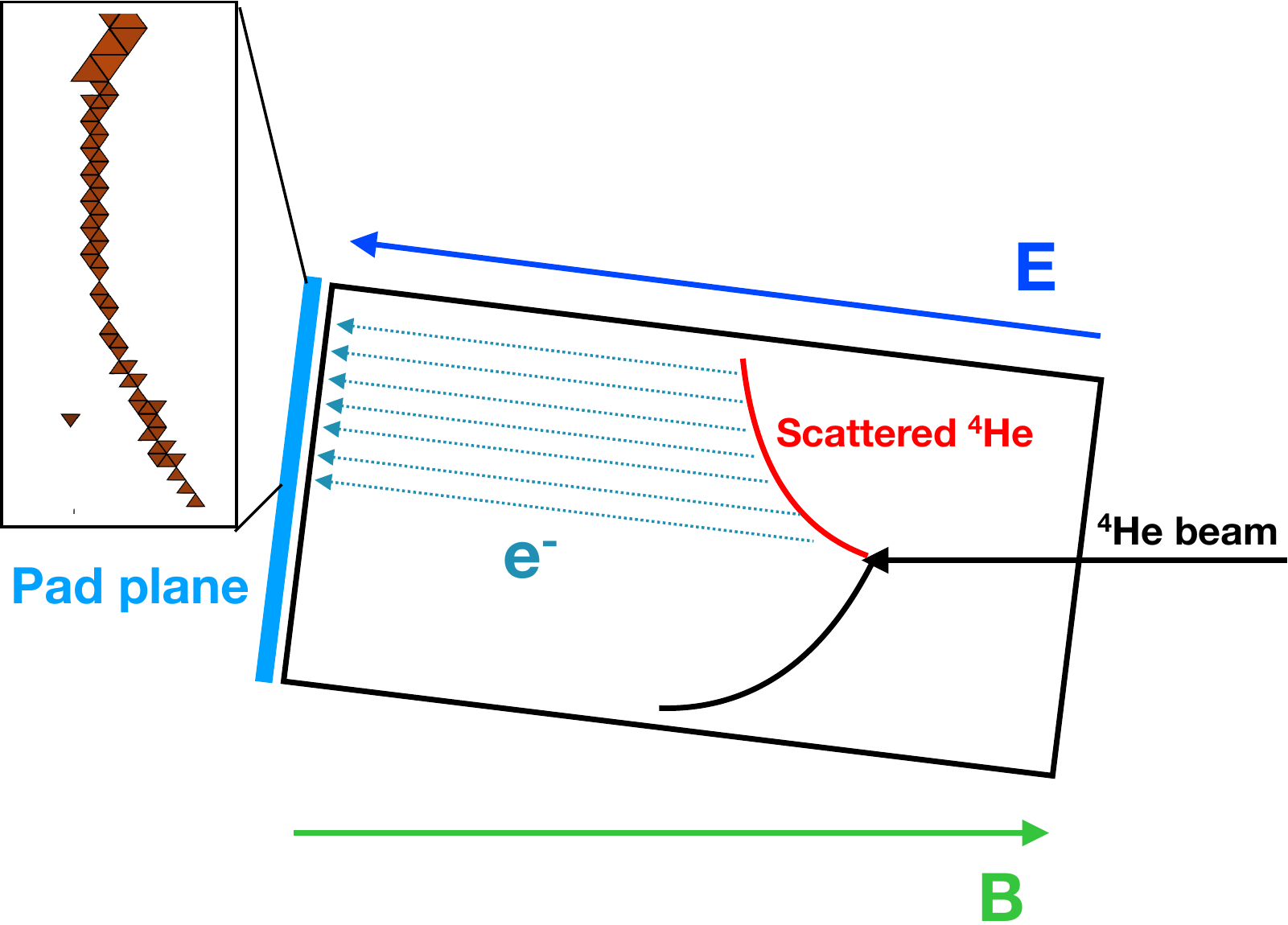}\label{fig:attpc_sketch_a}}
\subfigure[Hit pattern visualized with the AT-TPC analysis software]{\includegraphics[width=0.90\columnwidth]{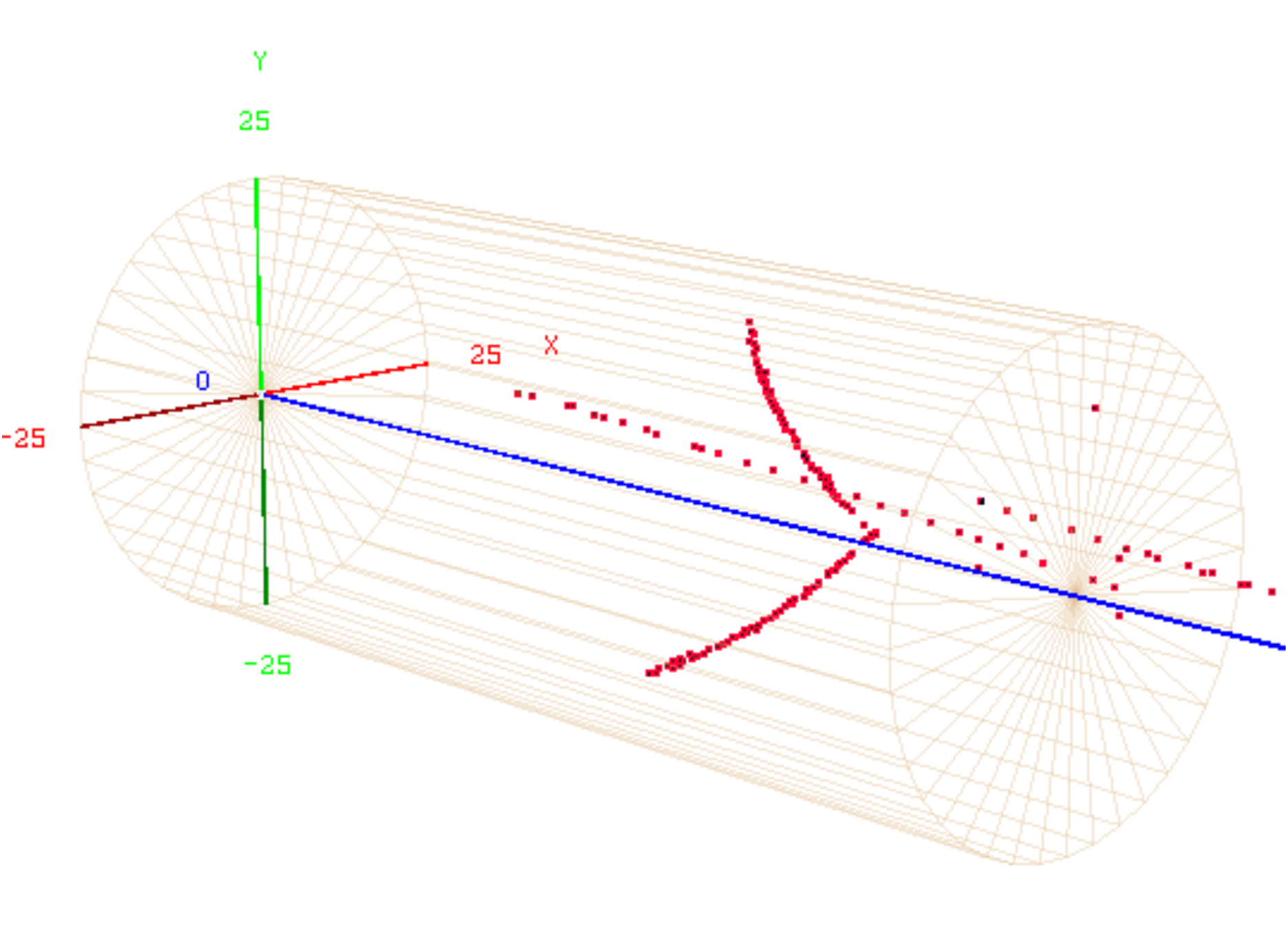}\label{fig:attpc_sketch_b}}
\caption{\label{fig:attpc_sketch}A sketch of the AT-TPC representing a $^{4}$He+$^{4}$He event and a 3-dimensional hit pattern. A strong electric field (E) is applied within the gas volume to drift the electrons (e$^{-}$) to the pad plane where the tracks are projected on a triangular pad structure. Operating the detector in a magnetic field (B) allows the measurement of the curvature of the track.}
\end{figure}

\begin{figure}[t!]
\center
\subfigure[$^{4}$He+$^{4}$He event]{\includegraphics[width=0.7\columnwidth]{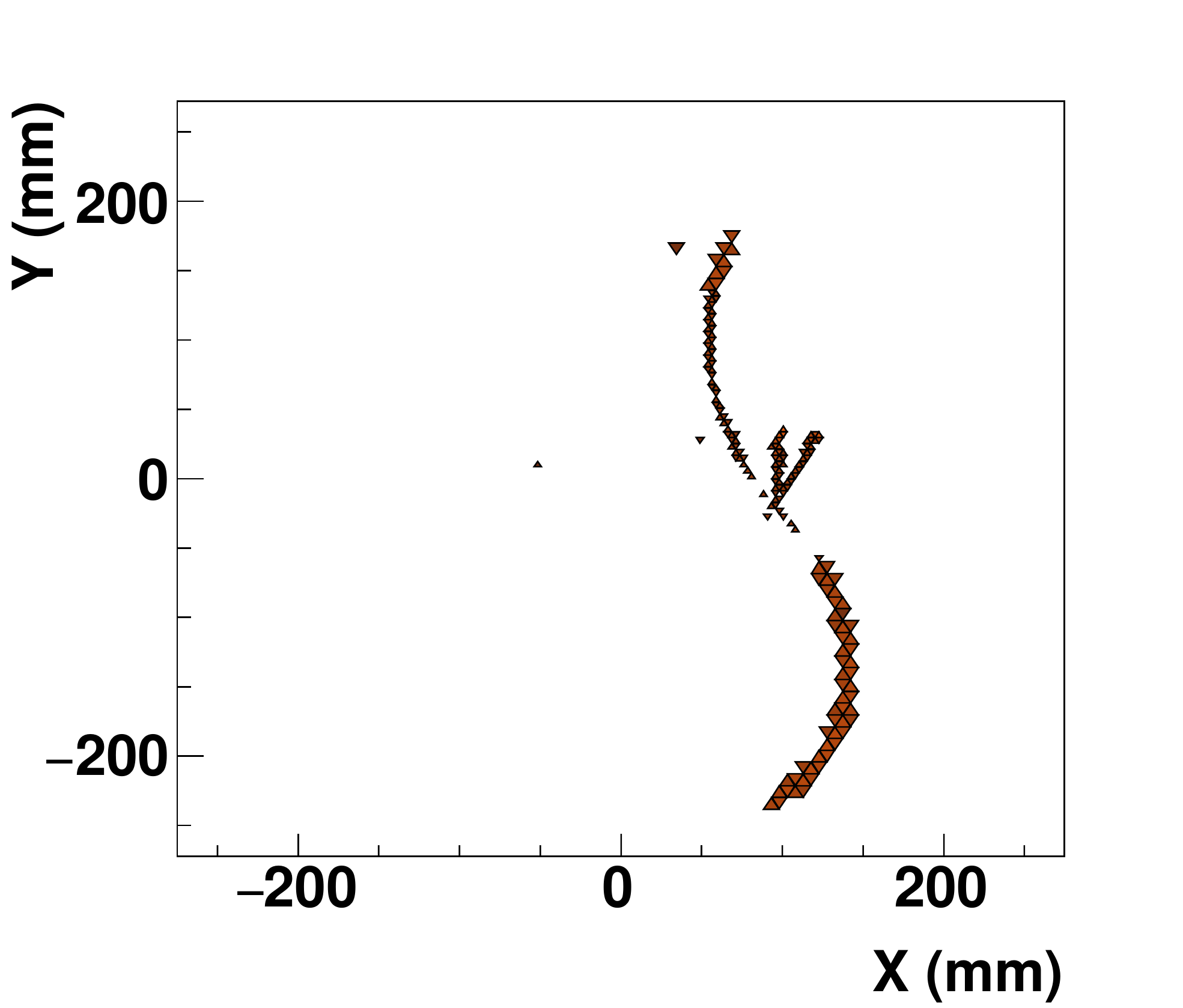}\label{fig:pad_plane_helium}}
\subfigure[$^{46}$Ar+p event]{\includegraphics[width=0.7\columnwidth]{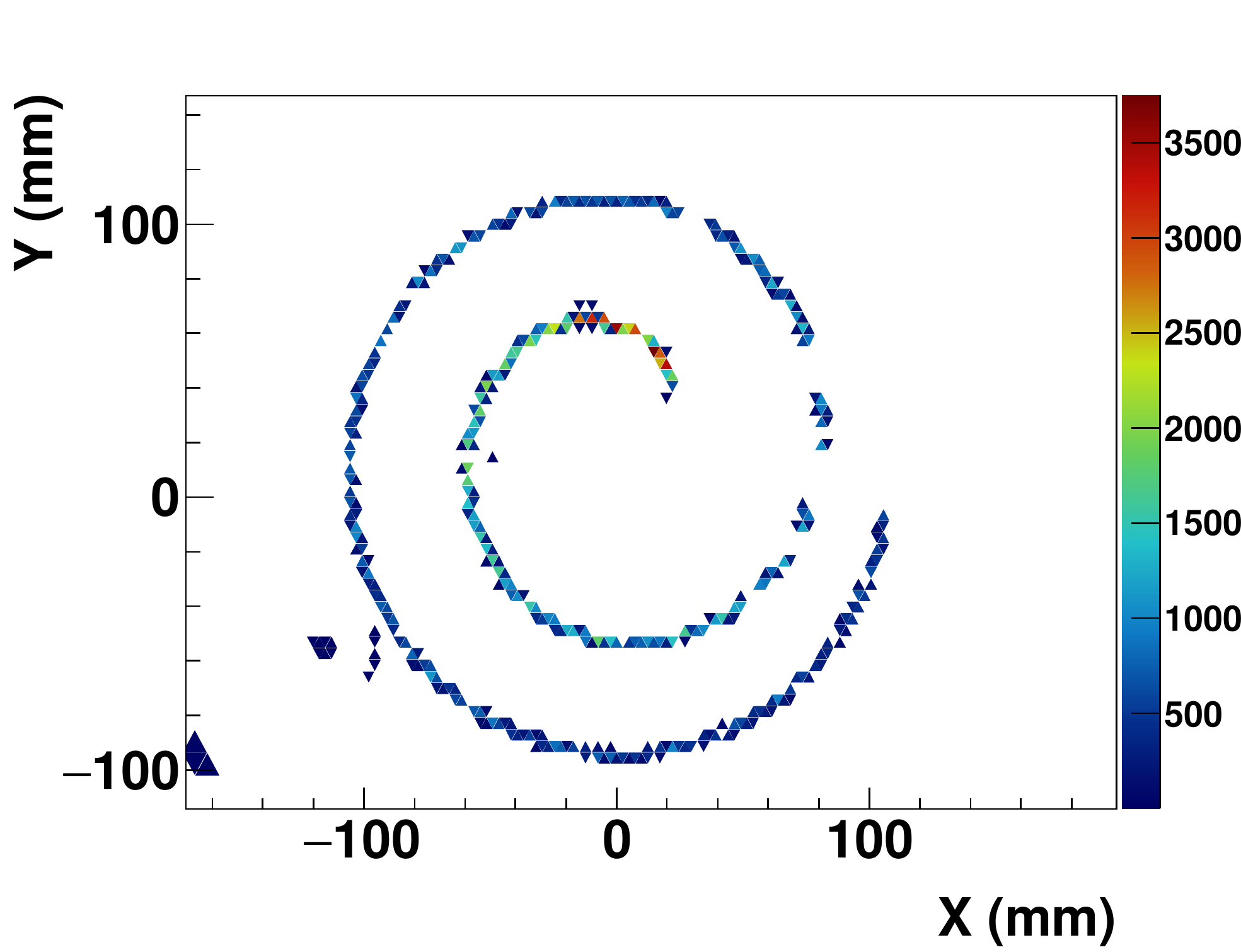}\label{fig:pad_plane_proton}}
\caption{\label{fig:pad_plane}Projections on the pad plane of two different events. The color scale shows the activation level or charge collected in each pad. As the particle stops, the radius of curvature decreases and the energy loss increases.}
\end{figure}

An sketch of the AT-TPC and a 3-dimensional hit pattern of a $^{4}$He+$^{4}$He reaction event are shown in Fig.~\ref{fig:attpc_sketch}.  In this event, a $^{4}$He beam particle penetrates the detector from the right side, and interacts with a $^{4}$He atom of the gas. After the reaction, both particles are emitted in forward direction, with a characteristic angle and energy, describing curved trajectories. Both particles stop in the gas as they continuously lose energy. Each point of the hit pattern corresponds to a pad that recorded the ionization electrons (e$^{-}$) that drifted from the point where the particle ionized the gas to the pad plane (situated on the left end cap). Fig.~\ref{fig:pad_plane_helium} shows the projection of the ionization electrons into the pads of the plane for this $^{4}$He+$^{4}$He event. In order to remove noise points, the resulting hit pattern was filtered using the Statistical Outlier Removal filter provided by the Point Cloud Library (PCL)~\cite{Rusu11}. It is worth mentioning that due to the stochastic nature of the time structure of the beam, two additional (pile-up) $^{4}$He beam particles entered the detector within the same time window as the particle that generated the event. In order to mitigate the effect of pile-up, which may cause a misidentification of the beam particle that reacted, the detector was tilted around 7$^{o}$~\cite{BRADT201765}. As can be seen in Fig.~\ref{fig:pad_plane_helium}, the pile-up particles are projected into two short straight lines emerging from the same vertex. In addition, the track of the beam particle that generated the event does not appear in the hit pattern, probably because the reaction happened very close to the entrance window. Note that the absolute $z$ coordinate depends on the timing of the particle that is detected first.

Fig.~\ref{fig:pad_plane_proton} shows the pad plane image of a proton emitted in the $^{46}$Ar+p reaction~\cite{Bradt2018}. In this case, the gain in the region of the pad plane where the $^{46}$Ar was projected was highly reduced to avoid saturation effects, and therefore the corresponding track is not shown. As can be clearly seen, proton tracks have a stronger curvature that changes as the particle slows down. The complexity for this case resides in the fact that the phase space (i.e.: number of possible angle-energy states) of the proton can be vast, depending on the experiment. Since the length and curvature of the spiral strongly depends on the energy of the particle, it is extremely challenging to find a unique solution based on parametric approaches, as these curves cannot be described by a simple analytical approach.

According to these two scenarios that represent typical experiments with ATs, the track finding algorithm proposed in this work has to cope with several difficulties: multiple tracks have to be found and categorized to provide an efficient pile-up rejection mechanism, such tracks have to be found without any prior knowledge of their shape, the vertex has to be found even in absence of the beam particle track, the algorithm has to be immune to the large energy dynamic range for different particles and to noise sources (electronics, detector discharge...). The algorithm must be also capable of reconstructing a {\it non-continuous} track that has multiple gaps in the hit pattern, as the ones shown in Fig.~\ref{fig:pad_plane}. In addition, taking into account the amount of data generated during an experiment, the algorithm must be computationally inexpensive, or in the best-case scenario, run concurrently. While tracking every particle is desirable, the reconstruction of the kinematics of the reaction requires only the inference of the energy and angle of one of the reaction products of a binary reaction, preferably the lighter one. This entails a considerable flexibility in the design of the algorithm that can be adapted to different situations.

\begin{table*}[t]
\centering
\caption{\label{tbl:steps-params} Overview over the algorithm and the external parameters controlling each step. Default values are recommendations taken from section \ref{sec:evaluation}.}
\smallskip
\begin{tabular}{|l|l|l|}
  \hline
  {\em Step} & {\em Parameter} & {\em Default value} \\
  \hline
  1) neighborship smoothing & $r_{\mbox{\scriptsize\it smooth}}$ = neighbor distance & $2\cdot d_{\mbox{\scriptsize\it NN}}$\\
  \hline
  2) triplet building & $k_{\mbox{\scriptsize\it triplet}}$ = tested neighbors of triplet mid point & $19$ \\
  & $n_{\mbox{\scriptsize\it triplet}}$ = max number of triplets to one mid point & $3$ \\
  & $a_{\mbox{\scriptsize\it triplet}} = 1-\cos\alpha$, where $\alpha$ is & $0.03$ \\
  & \hspace{2em}the angle between the two triplet branches & \\
  \hline
  3) triplet clustering & $s_{\mbox{\scriptsize\it cluster}}$ = distance scale factor in metric & $d_{\mbox{\scriptsize\it NN}}/3$ \\
  & $t_{\mbox{\scriptsize\it cluster}}$ = threshold for {\em cdist} in clustering & $3.5$ \\
  \hline
  4) pruning & $m_{\mbox{\scriptsize\it cluster}}$ = min number of triplets per cluster & $8$ \\
  \hline
\end{tabular}
\end{table*}

In order to test the performance of the proposed algorithm, we have prepared three different sets of data extracted from the above-mentioned experiments: two sets extracted from the $^{4}$He+$^{4}$He experiment with (data set I) and without magnetic field (data set III), respectively, and a set consisting of $^{46}$Ar+p events with magnetic field (data set II). Each data set is a collection of events acquired during these experiments. The $^{4}$He+$^{4}$He data sets are characterized, as explained before, by two tracks of different length, curved or straight depending on the magnetic field,  and an additional straight track corresponding to the beam particle. These three tracks are connected by a common point known as reaction vertex. In the $^{46}$Ar+p data, only protons tracks, characterized by a very different radius of curvature, conform the hit pattern.

\section{Track detection algorithm}
\label{sec:algorithm}
The algorithm for grouping the points into possibly overlapping clusters or noise consists of four steps:
\begin{enumerate}
\item smoothing by position averaging of neighboring points
\item finding triplets of approximately collinear points
\item single link hierarchical clustering of the triplets
\item pruning by removal of small clusters
\end{enumerate}

Each step can be controlled by external parameters which are listed in table \ref{tbl:steps-params}. The parameter values should be adjusted for each experiments. The default values are only meant as first guesses, when no other information is available.

The individual steps and the meaning of the parameters are described in detail in the following subsections. It should be noted that, although steps 2)-4) are done on the smoothed data points, the point indices are not lost during smoothing, so that the final clustering actually is a grouping of the {\em original} points.

\subsection{Pre-processing}
\label{sec:algorithm:pre}
Some steps in the algorithm are controlled by thresholds on distances. Although these thresholds can be provided by an operator, it is preferable to have reasonable guesses for these thresholds based on internal properties of the data. One particular property is the distance between neighboring points. This distance varies with the physical properties of the particles (e.g. their velocity) and the ambient chamber gas (e.g. its pressure) and is thus different from experiment to experiment. From a geometrical point of view, it has the effect of a scale parameter and can thus be considered as a {\em characteristic length} for the experimental setup. It is also an indicator for the spread of the points around the actual particle track.

\begin{figure}[t]
  \centering
  \includegraphics[width=1.0\columnwidth]{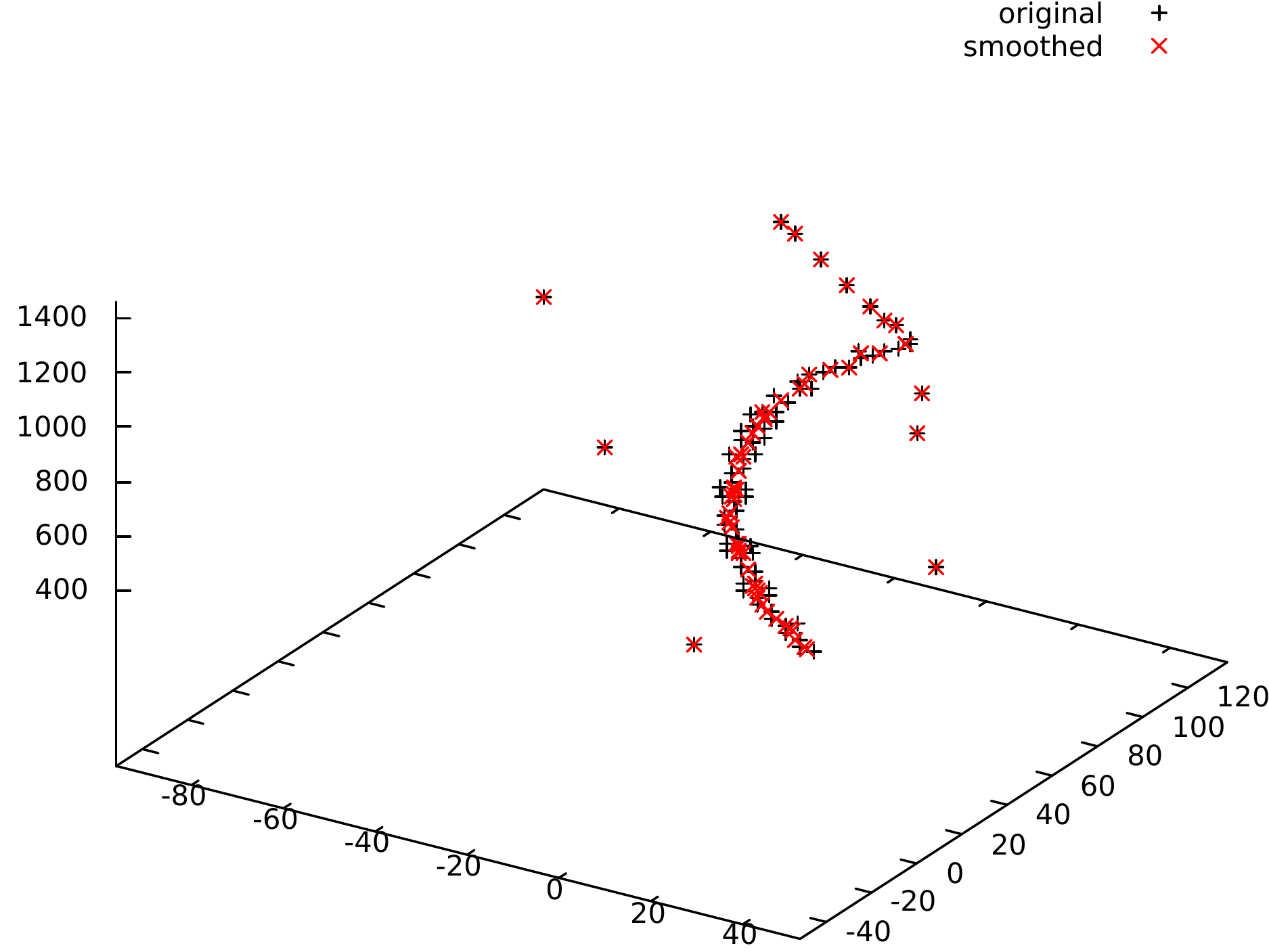}
\caption{\label{fig:nnsmoothing}Effect of the neighborhood smoothing with $r_{\mbox{\scriptsize\it smooth}}=2\cdot d_{\mbox{\scriptsize\it NN}}$.}
\end{figure}

To obtain an estimator for the point distance, we compute, for each point, the distance to its nearest neighbors. The first quartile of all these distances is our characteristic length $d_{\mbox{\scriptsize\it NN}}$. We have chosen the first quartile, because it typically lies within a track of a slower particle, even in the presence of considerable noise. This value will vary from point cloud to point cloud even for the same experimental setup, but it nevertheless can be considered as a scale parameter, because when all coordinates are scaled with the same factor, $d_{\mbox{\scriptsize\it NN}}$ will scale with this factor, too.

To reduce the spread of the measured points around the particle trajectories, we perform a {\em neighborship smoothing}, which replaces the coordinates of each point $\vec{p}$ by the mean $\sum_{i=1}^k \vec{q}_i /k$ of the points $\vec{q}_1,\ldots,\vec{q}_k$ in its neighborship. A point $\vec{q}_i$ is considered to belong to the neighborship of $\vec{p}$, if its distance $\|\vec{q}_i-\vec{p}\|$ is less than a threshold $r_{\mbox{\scriptsize\it smooth}}$. The averaging sum always contains at least one point, because a point is always part of its own neighborship. As $d_{\mbox{\scriptsize\it NN}}$ is a measure for the spread of the points, we use it as a scale factor for the default value, i.e., $r_{\mbox{\scriptsize\it smooth}}= 2\cdot d_{\mbox{\scriptsize\it NN}}$. Figure \ref{fig:nnsmoothing} shows the effect of this smoothing operation: points without neighbors in the $r_{\mbox{\scriptsize\it smooth}}$ range remain unchanged, whereas points in denser regions are moved towards the middle axis of the track.

\subsection{Triplet grouping}
\label{sec:algorithm:triplets}
\begin{figure}[t]
  \centering
  \subfigure[nodes $A,B,C$ and angle $\alpha$]{\includegraphics[width=0.6\columnwidth]{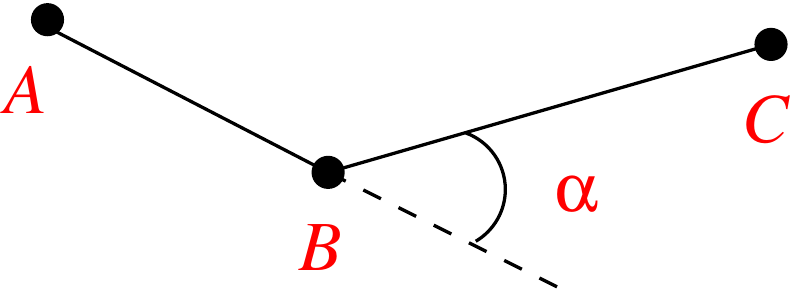}\label{fig:triplet:a}}\\
  \subfigure[midpoint $\vec{m}$ and direction $\vec{e}$]{\includegraphics[width=0.6\columnwidth]{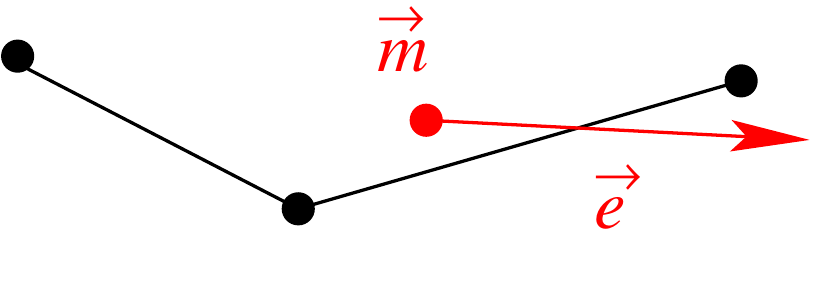}\label{fig:triplet:b}}
\caption{\label{fig:triplet}Properties of a triplet.}
\end{figure}

The second step in the algorithm consists of building groups of three approximately collinear points, i.e., {\em triplets}. An example of a triplet is shown in figure \ref{fig:triplet:a}. As TPC data are ordered in time direction, the indices of the node points $A=\vec{q_i}$, $B=\vec{q_j}$, and $C=\vec{q}_k$ are subject to the condition $i<j<k$. The cosine of the angle $\alpha$ between the triplet branches is given by
\begin{equation}
  \label{eq:triplet-alpha}
  \cos(\alpha) = \frac{\langle \overline{AB},\overline{BC}\rangle}{\|\overline{AB}\|\cdot\|\overline{BC}\|} =
  \frac{\langle \vec{q}_j-\vec{q}_i, \vec{q}_k-\vec{q}_j\rangle}{\|\vec{q}_j-\vec{q}_i\|\cdot\|\vec{q}_k-\vec{q}_j\|}
\end{equation}
For the hierarchical clustering described in the following subsection, each triplet is represented by two vectors, the midpoint $\vec{m}$ and the direction $\vec{e}$ between the outer points:
\begin{equation}
  \label{eq:triplet-me}
  \vec{m}=\frac{1}{3}\left( \vec{q}_i + \vec{q}_j + \vec{q}_k\right)
  \quad\mbox{and}\quad
  \vec{e}=\frac{\vec{q}_k-\vec{q}_i}{\|\vec{q}_k-\vec{q}_i\|}
\end{equation}

The triplets are computed from the point cloud in a loop over all points. Each point is considered as a possible midpoint $B$, and all points among its $k_{\mbox{\scriptsize\it triplet}}$ nearest neighbors are tried as candidate points $A$ and $B$. The triplet is discarded, if the triplet angle $\alpha)$ is greater than the threshold implied by $a_{\mbox{\scriptsize\it triplet}}$, or, equivalently, if
\begin{equation}
  \frac{\langle \overline{AB},\overline{BC}\rangle}{\|\overline{AB}\|\cdot\|\overline{BC}\|} < 1-a_{\mbox{\scriptsize\it triplet}}
\end{equation}
From the remaining triplets of each midpoint, only the $n_{\mbox{\scriptsize\it triplet}}$ triplets with the smallest angle $\alpha$, or, equivalently, with the greatest $\cos(\alpha)$ are kept. For $n$ points, the total number of triplets is thus not more than $n\cdot n_{\mbox{\scriptsize\it triplet}}$.

\subsection{Hierarchical clustering}
\label{sec:algorithm:clustering}
The third step of the algorithm consists in a hierarchical clustering, which starts with each triplet as a separate cluster and merges in each iteration two clusters. The exact procedure is described in Algorithm \ref{alg:clustering}. It depends on a distance measure $\mbox{\it cdist}(C_i,C_j)$ on clusters $C_i$ that can be constructed from a distance measure $d(x,y)$ on triplets $x,y$ in different ways, known as {\em complete link}, {\em average link}, or {\em single link} clustering \cite{theodoridis09}. In our situation, only the single link method is appropriate because, for curved tracks, the angle distance between triplets of the same cluster can become large. The single link method defines the cluster distance as
\begin{equation}
  \mbox{\it cdist}(C_i,C_j) = \min\{d(x,y) \mid x\in C_i, y\in C_j\}
\end{equation}

\begin{algorithm}[t!]
\caption{\label{alg:clustering}Hierarchical clustering}
\begin{algorithmic}[1]
\Require set of triplets $X = \{x_1,\ldots,x_m\}$, threshold $t_{\mbox{\scriptsize\it cluster}}$ on cluster distance
\Ensure triplet clustering $M=\{C_1,\ldots,C_k\}$
\State $M \gets \{C_i=\{ x_i \}, i=1,\ldots,m \}$
\For{$i=1,\ldots,m-1$}
\State \label{alg:custering:pairs}from all pairs $C_i, C_j \in A$,
select the pair with smallest distance $\mbox{\it cdist}(C_i, C_j)$
\If {$\mbox{\it cdist}(C_i, C_j) >t_{\mbox{\scriptsize\it cluster}}$}
\State {\bf break}
\EndIf
\State $C_h\gets C_i\cup C_j$
\State $M \gets \left(M \setminus \{ C_i,C_j\}\right) \cup \{C_h\}$
\EndFor
\State \Return $M$
\end{algorithmic}
\end{algorithm}

This leads to the question how to define a metric on triplets. Let us start with the observation that two triplets $(A_i,B_i,C_i)$ and ($A_j,B_j,C_j)$ are similar when three conditions hold:
\begin{enumerate}
\item the perpendicular distance $d_1^\perp$ between the midpoint $\vec{m_i}$ and the extrapolated line $\vec{m_j}+\lambda \vec{e_j}$ is small, which is
  \begin{equation}
    d_1^\perp = \| \vec{m_j} - \vec{m_i} + \langle\vec{m_i}-\vec{m_j}, \vec{e_j}\rangle\cdot\vec{e_j} \|
  \end{equation}
\item the perpendicular distance $d_2^\perp$ between the midpoint $\vec{m_j}$ and the extrapolated line $\vec{m_i}+\lambda \vec{e_i}$ is small, which is
  \begin{equation}
    d_2^\perp = \| \vec{m_i} - \vec{m_j} + \langle\vec{m_j}-\vec{m_i}, \vec{e_i}\rangle\cdot\vec{e_i} \|
  \end{equation}
\item the angle $\varphi$ between their direction vectors $\vec{e}$ is small, which is
  \begin{equation}
    \varphi = \cos^{-1}\left(|\langle\vec{e_i},\vec{e_j}\rangle|\right)
  \end{equation}
\end{enumerate}
We combine these three distance measures into a single measure via
\begin{equation}
  \label{eq:tripletdistance}
  d\Big((A_i,B_i,C_i), (A_j,B_j,C_j)\Big) = \frac{\max\{d_1^\perp,d_2^\perp\}}{s_{\mbox{\scriptsize\it cluster}}} + |\tan\varphi|
\end{equation}
We use the tangens as an angle distance measure, and not one minus the cosine, because the tangens goes to infinity as $\varphi\to\pm\pi/2$, which means that perpendicular triplets have an infinite distance, regardless of their spatial distance. The scale factor $s_{\mbox{\scriptsize\it cluster}}$ allows for controlling the relative effect of angle and perpendicular distance.

\begin{figure}[t]
  \centering
  \includegraphics[width=0.8\columnwidth]{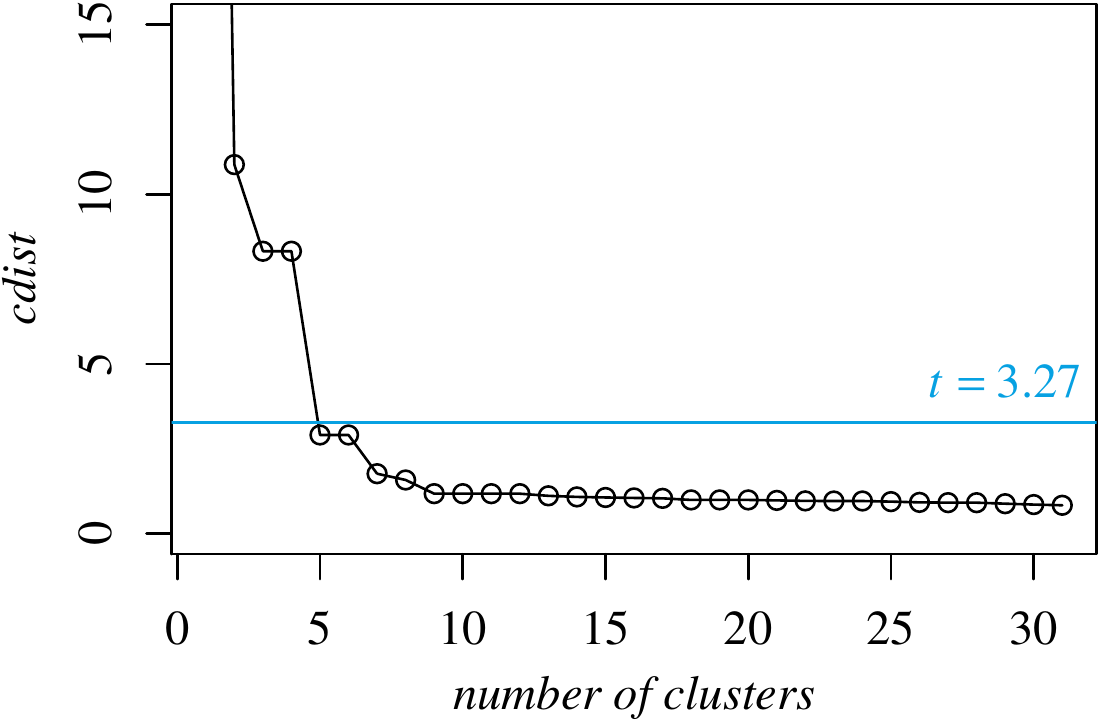}
\caption{\label{fig:cdist}In this example, the threshold $t_{\mbox{\scriptsize\it cluster}}=3.27$ stops the clustering such that five clusters are obtained.}
\end{figure}

We recommend to set the default value for $s_{\mbox{\scriptsize\it cluster}}$ proportional to $d_{\mbox{\scriptsize\it NN}}$. This has the effect that the distance measure becomes scale invariant, because when all point coordinates are scaled with the same factor, $d_{\mbox{\scriptsize\it NN}}$ scales with this factor too. A scale invariant distance measure has the advantage, that an absolute threshold $t_{\mbox{\scriptsize\it cluster}}$ can be used for stopping the clustering process in Algorithm \ref{alg:clustering}, because the distances fall into an approximately predictable range. Figure \ref{fig:cdist} shows a typical example, for which the threshold $t_{\mbox{\scriptsize\it cluster}}$ stops the clustering at the ``elbow'' of the cluster distances {\em cdist}.

\subsection{Pruning}
\label{sec:algorithm:pruning}
To distinguish actual tracks from random clusters due to noise, it is necessary to remove some of the clusters in a post-processing step. We have applied a very simple rule and removed all clusters containing less than $m_{\mbox{\scriptsize\it cluster}}$ triplets. If the data is known to be almost noise free, $m_{\mbox{\scriptsize\it cluster}}$ can be set to two. As this is rarely the case, we recommend a default value of $m_{\mbox{\scriptsize\it cluster}}=3$.

\subsection{Runtime considerations}
\label{sec:algorithm:runtime}
As the four steps are processed sequentially, the order of the total runtime is the order of slowest step. Here is a runtime analysis of the individual steps where $n$ is the number of points in the point cloud:
\begin{enumerate}
\item Both the estimation of $d_{\mbox{\it NN}}$ and the neighborship smoothing are $O(n^2)$ operations when they are implemented by simple loops. The nearest neighbor retrieval can be reduced to $O(n\log n)$ time with a kd-tree, but only on average. To achieve worst case $O(n\log n)$ runtime, other algorithms have been suggested \cite{beygelzimer06,vaidya89}. For the distance search, a utilization of a kd-tree leads to $O(n\cdot n^{2/3})$ runtime, but this can be further reduced to $O(n\log^3 n)$ when a range tree is used \cite{berg00}.
\item The number of triplets that need to be evaluated for building the triplets is ${n \choose 3}$, which is $O(n^3)$. Utilization of the point order, as suggested in section \ref{sec:algorithm:triplets}, reduces the runtime by a constant factor, but does not change the exponent. As we do not try all possible triplets, but only search for candidate points among the $k_{\mbox{\scriptsize\it triplet}}$ nearest neighbors, the runtime reduces to $O(k_{\mbox{\scriptsize\it triplet}}^3\cdot n\log n)$, where the factor $n\log n$ stems from the all-nearest-neighbor search.
\item The runtime complexity of Algorithm \ref{alg:clustering} is $O(m^3)$ where $m$ is the number of triplets \cite{theodoridis09}, but this can be reduced to $O(m^2)$ with the utilization of Rohlf's MST-algorithm \cite{muellner11,rohlf73}. As we make the restriction of keeping only the $n_{\mbox{\scriptsize\it triplet}}$ ``best'' triplets for each candidate midpoint, the number of triplets is only $O(n)$, not $O(n^3)$, as it were if we kept all triplets below the given collinearity threshold $a_{\mbox{\scriptsize\it triplet}}$. The runtime of the clustering step is thus $O(n^2)$.
\item As we cannot have more clusters than triplets, the runtime of the pruning is $O(m)$, which is $O(n)$ due to the restriction mentioned in the preceding step.
\end{enumerate}
The runtime is thus dominated by the hierarchical clustering and is $O(n^2)$ in total. For the nearest neighbor and range search, we have used the kd-tree implementation that came with the Point Cloud Library (PCL) \cite{Rusu11}. For the hierarchical clustering, we have used the library {\em fastcluster} \cite{muellner13}. The worst-case runtime occurred on a point cloud consisting of 666 points with much noise and strongly curved tracks, which resulted in a large number of candidate triplets. In this case, the runtime was about 0.5 seconds on an Intel Core i5-6500 CPU @ 3.20GHz processor. The clustering step took about 70\% of the total runtime.

\section{Evaluation}
\label{sec:evaluation}
With an evaluation of our algorithm, we pursue two different aims. One is the determination of parameters that lead to a decent clustering of the points into particle tracks. The other one is an assessment how accurate the physical properties of of the trajectories can be detected. To this end, we need ground truth data, in which each point is labeled with its track membership, and a similarity measure between the ground truth data and a test clustering.

It should be noted that, although both the parameter optimization and the quality assessment are done on the same test data, we did not optimize the parameters for each point cloud, but used a global optimum instead. That way, each individual point cloud only has a small effect on the parameter choice, and the quality assessment is not too much optimistically biased.

\begin{figure}[t!]
  \centering
  \includegraphics[width=1.0\columnwidth]{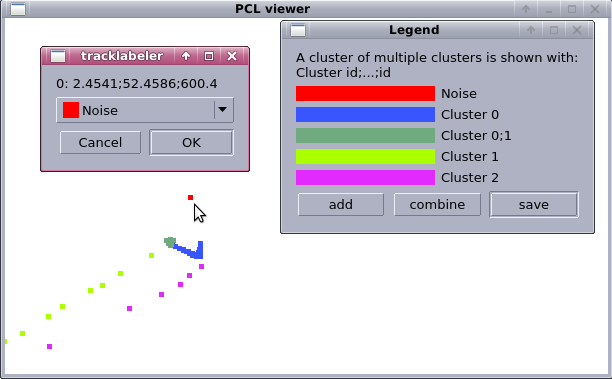}
\caption{\label{fig:tracklabeler}Screen-shot of the interactive software for ground truth labeling.}
\end{figure}

\subsection{Ground truth data}
\label{sec:evaluation:groundtruth}
For the creation of ground truth data, we have written an interactive software that displays the point cloud in such a way that cluster membership is visualized by colorization. Each point is assigned to exactly one cluster or it is labeled as noise. The list of clusters can be edited, and the cluster label of a point can be changed through a popup menu that appears when clicking on the point (see Figure \ref{fig:tracklabeler}). The point coordinates together with their cluster labels are eventually stored in a CSV file.

Obviously, this approach does not allow for vertex points to simultaneously belong to more than one trajectory. We have therefore covered this case by introducing special clusters that represent multiple memberships, e.g., ``cluster 2;4'' as a cluster label for points that belong both to clusters labeled as ``2'' and ``4''. Although this could theoretically result in up to additional $k(k-1)/2$ clusters, where $k$ is the number of particle tracks, in practice there are only few crossing trajectories and consequently only few vertices. It is thus a both simple and feasible way of encoding ambiguities.

\begin{table}[t]
\centering
\caption{\label{tbl:gt-stats} Some statistics of the ground truth labels in different data sets. ``\#'' stands for ``number of''.}
\smallskip
\begin{tabular}{|l|rrr|}
  \hline
  & \multicolumn{3}{|c|}{Data set} \\
  {\em Property} & \multicolumn{1}{c}{I} & \multicolumn{1}{c}{II} & \multicolumn{1}{c|}{III}  \\ \hline
  \# point clouds & 99 & 136 & 104 \\
  \# points & 8 472 & 31 724 & 19 274\\
  noise fraction & 6.1\% & 27.7\% & 3.3\% \\ \hline
  minimum \# tracks & 1 & 1 & 2 \\
  median \# tracks & 1 & 1 & 3 \\
  maximum \# tracks & 6 & 6 & 6 \\ \hline
  \# vertices & 37 & 0 & 103 \\ \hline
\end{tabular}
\end{table}

\subsection{Evaluation criteria}
\label{sec:evaluation:criteria}
We evaluate the quality of our algorithm with two different means: a generic measure for the distance between two different clusterings based on pairs of points, and by some specific physical properties of the detected clusters. The generic measure was used for optimizing the algorithm parameters, and the physical properties were evaluated to obtain an idea how good the algorithm works for processing AT measurements.

\subsubsection{Clustering evaluation metrics}
When comparing a test clustering with a ground truth clustering, the problem occurs that no correspondences between the cluster labels are known, and that such a mapping is not even well-defined because the number of clusters can be different, clusters may be split, merged or overlap. An established workaround for this problem is to consider all {\em pairs of points} and to check whether both points belong to the same (S) or to different (D) clusters in a clustering, which results, when comparing two clusterings, in four possible combinations (SS, SD, DS, or DD) \cite{pfitzner08,amigo09}. Over all pairs, the following numbers are counted:
\begin{itemize}
\item $a$ = number of pairs with SS
\item $b$ = number of pairs with SD
\item $c$ = number of pairs with DS
\item $d$ = number of pairs with DD
\end{itemize}
Actually, these are only three independent numbers, because their sum is the total number of pairs. Depending on whether all four dependent or only three independent numbers are used, the following similarity measures can be constructed:
\begin{eqnarray}
  \mbox{Rand statistic:}& & R=\frac{a+d}{a+b+c+d}\\
  \mbox{Jaccard coefficient:}& & J=\frac{a}{a+b+c}\quad\quad  
\end{eqnarray}
These coefficients fall into the range $0\leq R,J\leq 1$, where $R=1$ or $J=1$ only hold for exactly identical clusterings. Both coefficients have the desirable property that they decrease when a ground truth cluster is split up or when two ground truth clusters are merged in the test clustering \cite{amigo09}. It is always $R\geq J$ and, due to the typically large number $d$, the Rand coefficient often is much closer to one. The Jaccard coefficient is thus a more sensitive indicator for clustering similarity, and we have use $J$ for optimizing the parameters.

In our situation, we additionally have to deal with noise points, which we do by treating the noise cluster just like an ordinary cluster. Moreover, we take care of ambiguous cluster memberships at vertices in the following way: when one point of the pair is labeled as ambiguous in the ground truth data, then
\begin{itemize}
\item[a)] if the other point belongs to a compatible ground truth cluster, the pair is counted as SD, if exactly one of the test cluster labels is ``noise'', otherwise it is counted as SS
\item[b)] if the other point belongs to an incompatible ground truth cluster, the pair is counted as DS or DD, dependent on the test cluster labels
\end{itemize}
Here, a ``compatible ground truth cluster'' is either the same cluster, or one of the clusters that are represented by the ambiguous ``overlap cluster''. Condition a) results in a slightly optimistic bias of the similarity measures because, at vertices, other than noise labels in the test clustering are disregarded. This has, however, only a slight effect due to the small number of ambiguous points.

\subsubsection{Physical cluster properties}
For reconstructing the kinematics of an AT experiment, the following questions are of particular importance:
\begin{enumerate}
\item When there is a curved track, is it detected?
\item When there is a vertex, is it detected?
\item What was the smallest curve range that the algorithm detected?
\end{enumerate}
{\em Curved tracks} only occurred in test sets I \& II, so that the first question only applies to these data sets. We inspected all curved tracks in the ground truth data and counted how many of them were detected at all (more than 25\% of its point non-noise), how many were split into more than one cluster and how many were merged with a different cluster.

{\em Vertices} are junction points between different tracks. When they join only two tracks, they appear as kinks in a single curve, i.e., as a discontinuity of the velocity direction vector. In the ground truth data, points near vertices are labeled with multiple memberships. In the track detection algorithm, vertices lead to points that belong to more than one triplet which belong to different clusters. When a vertex joins more than two tracks, different cluster/triplet combinations are possible, and more than one vertex would be detected by the algorithm. In order to avoid this, we have merged close by vertices with an average link hierarchical clustering \cite{muellner13} based on their spatial distance with a cutoff distance of $2\cdot\overline{d_{\mbox{\scriptsize\it NN}}}$. For the remaining detected vertex points, we have measured two numbers: the fraction of groundtruth vertices that have been detected ({\em recall}), and the fraction of detected vertices that actually corresponded to groundtruth vertices ({\em precision}).

We define the {\em curve range} as the largest pairwise distance between all points in the same cluster. The smallest detected curve range was the smallest cluster that corresponded to a ground truth cluster.

\subsection{Results}
\label{sec:evaluation:results}
For finding optimal parameter combinations, we have used the Nelder-Mead optimization \cite{nelder65} as implemented in the {\em SciPy}\footnote{\url{https://scipy.org/}} function {\em scipy.optimize.minimize}. This algorithm only works with real valued parameters, but some of our parameters can only take integer values, namely the parameters $k_{\mbox{\scriptsize\it triplet}}$, $n_{\mbox{\scriptsize\it triplet}}$, and $m_{\mbox{\scriptsize\it cluster}}$. We therefore tested a fixed range of these integer parameters and optimized the remaining real valued parameters for each combination with {\em scipy.optimize.minimize}. The tested ranges for the integer parameters were $k_{\mbox{\scriptsize\it triplet}}\in\{10,\ldots,25\}$, $n_{\mbox{\scriptsize\it triplet}}\in\{2,\ldots,15\}$, and $m_{\mbox{\scriptsize\it cluster}}\in\{2,\ldots,20\}$.

\begin{table}[t]
\centering
\caption{\label{tbl:bestparams} Parameter values that lead to the best Jaccard coefficients on the different data sets. $\overline{d_{\mbox{\scriptsize\it NN}}}$ denotes the arithmetic mean of all $d_{\mbox{\scriptsize\it NN}}$ values in the specific data set.}
\smallskip
\begin{tabular}{|l|r|r|r|}
  \hline
   & {\em Set I} & {\em Set II} & {\em Set III}\\
  \hline
  best $J$ & 0.9726 & 0.8844 & 0.9706 \\\hline
  $r_{\mbox{\scriptsize\it smooth}}$ & 8.31 & 7.83 & 7.82 \\
  $k_{\mbox{\scriptsize\it triplet}}$ & 14 & 10 & 20 \\
  $n_{\mbox{\scriptsize\it triplet}}$ & 6 & 2 & 5 \\
  $a_{\mbox{\scriptsize\it triplet}}$ & 0.0291 & 0.0212 & 0.0330 \\
  $s_{\mbox{\scriptsize\it cluster}}$ & 0.9221 & 1.6159 & 1.0560 \\
  $t_{\mbox{\scriptsize\it cluster}}$ & 3.27 & 7.48 & 2.03 \\
  $m_{\mbox{\scriptsize\it cluster}}$ & 13 & 19 & 8 \\\hline
  $\overline{d_{\mbox{\scriptsize\it NN}}}$ & 4.7 & 3.3 & 3.4 \\
  \hline
\end{tabular}
\end{table}

Table \ref{tbl:bestparams} lists the parameters yielding the best Jaccard coefficient $J$. The values for $J$ are very high for data sets I and III, which shows that the algorithm works quite well on data with a small level of noise. On data set II, $J$ is smaller, yet still satisfactory when the poor data quality is taken into account. The noise level has impact on the parameters $m_{\mbox{\scriptsize\it cluster}}$ and $t_{\mbox{\scriptsize\it cluster}}$, too. This is easily understood, because more noise will lead to more random clusters, which need to be filtered out by a higher threshold for minimum number of triplets, $m_{\mbox{\scriptsize\it cluster}}$. A higher threshold $m_{\mbox{\scriptsize\it cluster}}$ also allows for more tolerance with respect to the maximum cluster distance $t_{\mbox{\scriptsize\it cluster}}$.

Without parameter optimization, some recommendations for an automatic parameter choice based on the characteristic point distance, $d_{\mbox{\scriptsize\it NN}}$, can also be drawn from table \ref{tbl:bestparams}. The smoothing radius can be set to $r_{\mbox{\scriptsize\it smooth}}\approx 2\cdot d_{\mbox{\scriptsize\it NN}}$. The other scale dependent parameter, the weighting factor for the spatial distance with respect to the angle distance in Eq.~(\ref{eq:tripletdistance}), can be set to $s_{\mbox{\scriptsize\it cluster}}\approx d_{\mbox{\scriptsize\it NN}}/3$.

The parameter $k_{\mbox{\scriptsize\it triplet}}$, which limits the number of neighbor points considered for building triplet, should be set according to the expected track curvature: the optimal value in table \ref{tbl:bestparams} is higher for data sets without (set III) or fewer (set I) bent trajectories.

\subsubsection{Data set I}
\begin{figure}[t]
  \centering
  \subfigure[\label{fig:set1-merge}merged tracks and missed vertex]{\includegraphics[width=0.5\columnwidth]{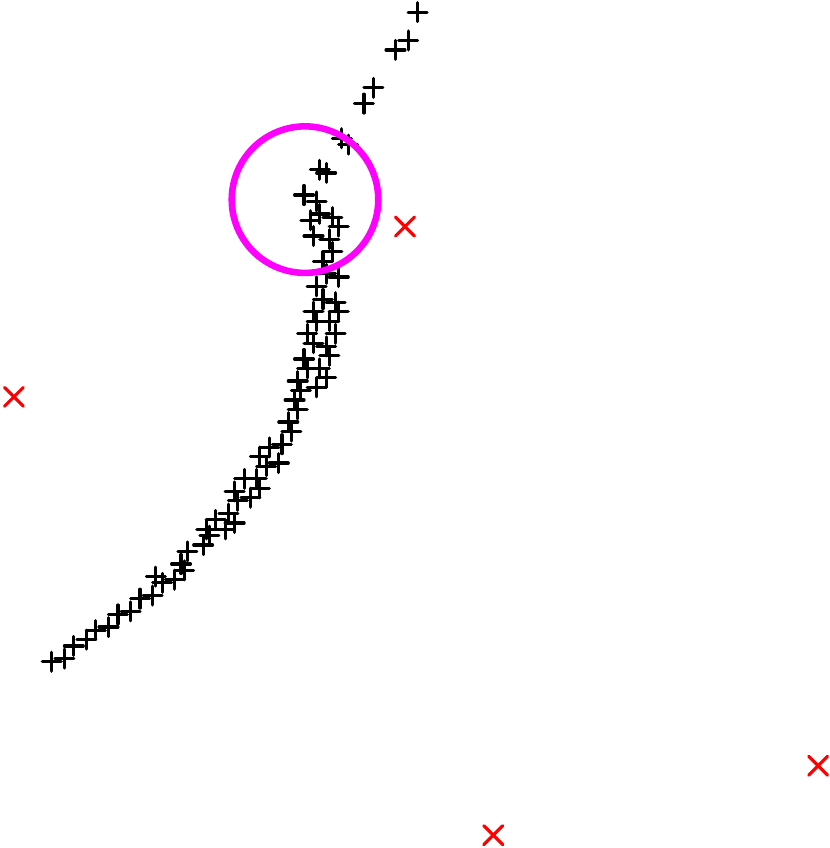}}
  \subfigure[\label{fig:set3-merge}two of three tracks merged, but vertex not missed]{\hspace{3em}\includegraphics[width=0.3\columnwidth]{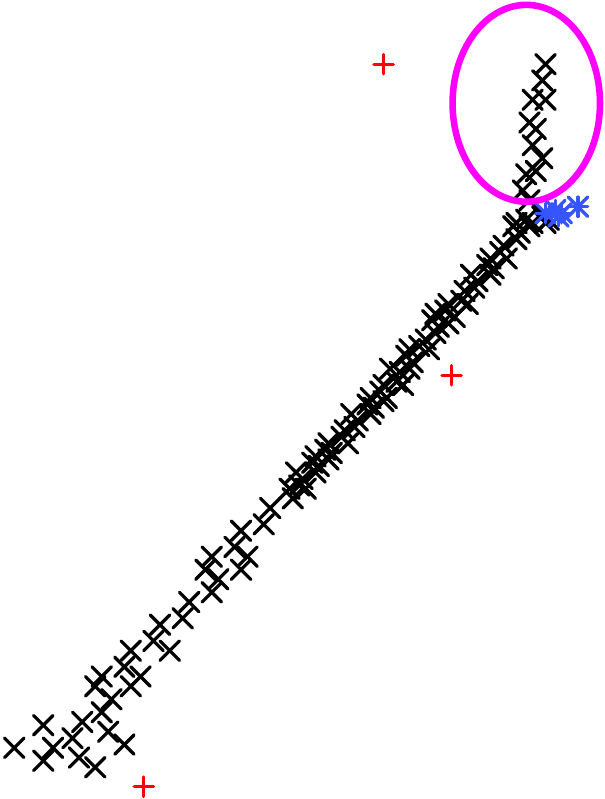}}
\caption{Examples for tracks merged into a single cluster. In the left example from set I, the track was not split at the vertex which lead to a miss of the marked vertex. In the right example from set III, the marked track was merged with the long track, but the vertex was nevertheless detected because of the third (blue) track. Red points have been classified as noise.}
\end{figure}

Out of the 99 curved tracks, 94 were correctly identified, none was split up, and 5 were merged with a touching straight track. An example for a merge is shown  in figure \ref{fig:set1-merge}. Except for one event, which had a straight track fragment with too few points, all other 4 merges could be correctly split up at the vertex when the parameters $k_{\mbox{\scriptsize\it triplet}}$ and $a_{\mbox{\scriptsize\it triplet}}$ were chosen smaller. As all of the merge cases returned only a single track, a practical approach would be to try more ``split-friendly'' parameters in cases for which the default parameters only yield a single track.

From a total of 37 vertices, 32 were correctly detected, which is a recall of 86\%. Only two non-existent vertices have been erroneously detected, which corresponds to a precision of 94\% among the returned vertices. The shortest range among the correctly identified tracks was about \SI{24}{mm}.

\subsubsection{Data set II}
Theoretically, all events in this data set only have one track in the form of a spiral plus noise. As this is known a priori from the experimental setup, a split up of a cluster by the algorithm does not pose much of a problem. We have therefore split up some spirals into more than one cluster in the ground truth data if the spiral had very wide gaps. This resulted in 217 tracks for the 136 events. Among these, 190 tracks were correctly detected, 18 were missed, 7 were split, and 2 were merged. Additionally, the algorithm returned 5 false positives, i.e., tracks that actually were noise.

\begin{figure}[t]
  \centering
  \subfigure[\label{fig:set2-error-false}noise classified as track segment]{\includegraphics[width=0.49\columnwidth]{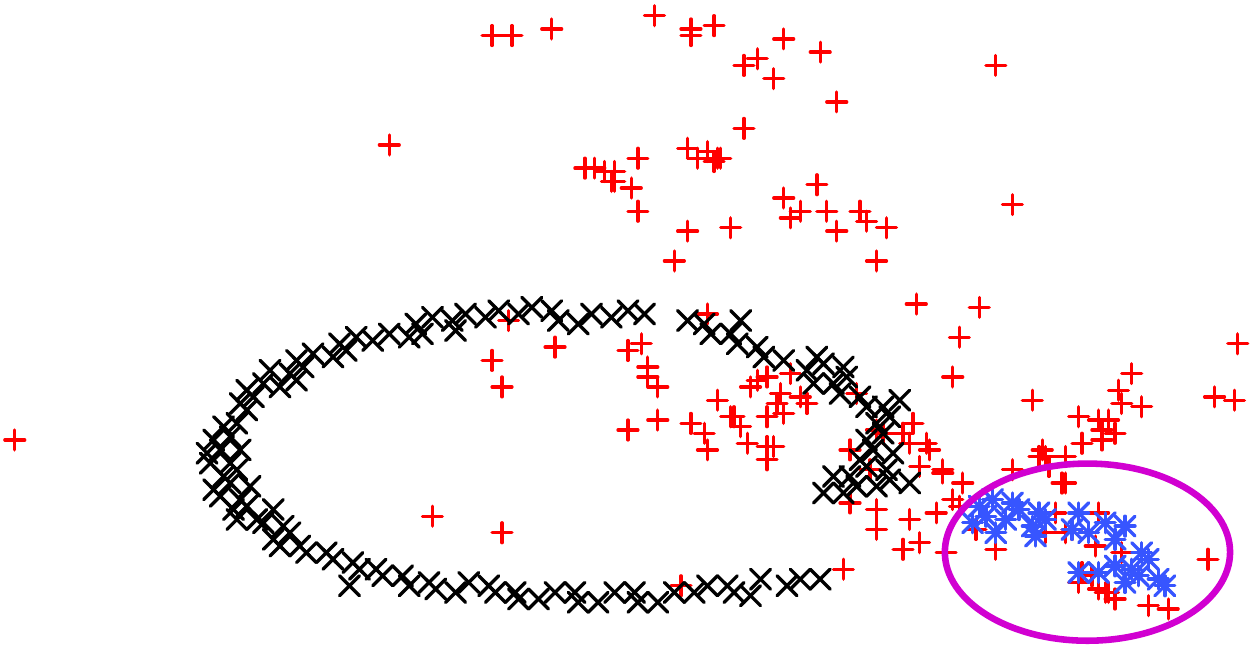}}
  \subfigure[\label{fig:set2-error-missed}missed track segment]{\includegraphics[width=0.49\columnwidth]{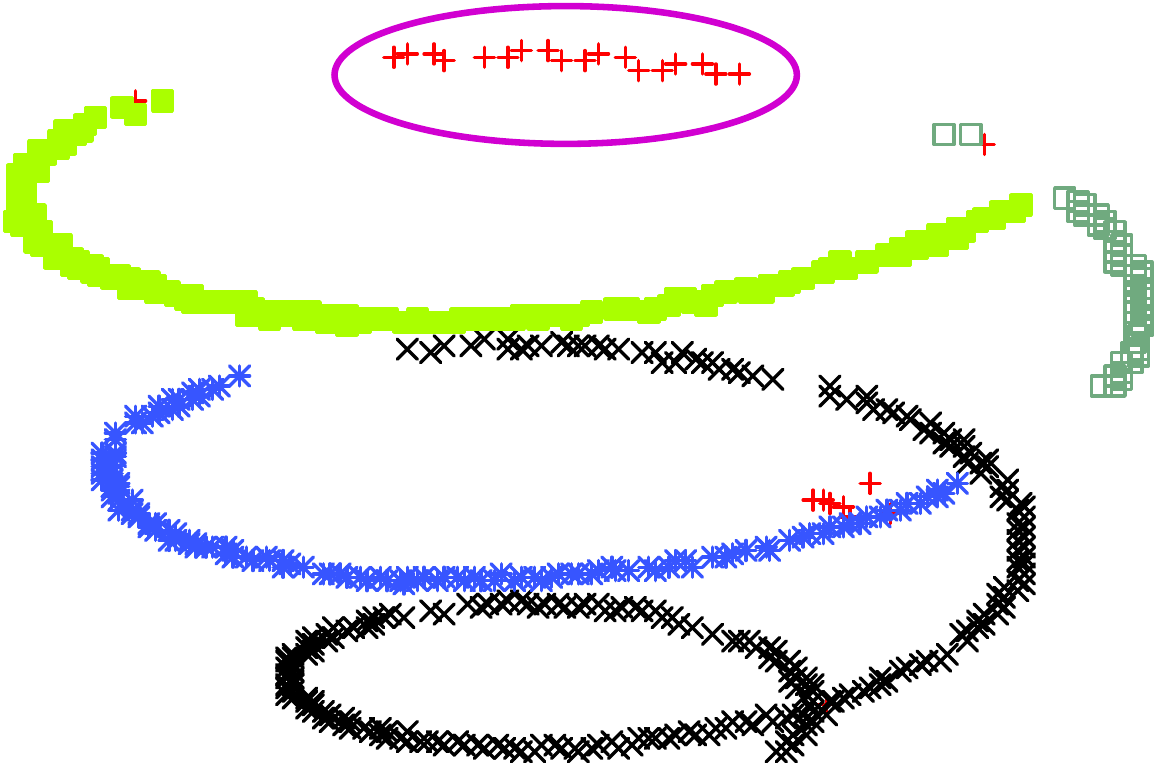}}
\caption{\label{fig:set2-error}Examples for track detection errors (marked) in data set II. Red points have been classified as noise.}
\end{figure}

Fig.~\ref{fig:set2-error} shows typical examples for these errors: in Fig.~\ref{fig:set2-error-false}, a track segment was found that actually was noise, and in Fig.~\ref{fig:set2-error-missed}, a track segment was missed because it contained less than $m_{\mbox{\scriptsize\it cluster}}$ triplets. Whilst the latter problem could be solved by decreasing $m_{\mbox{\scriptsize\it cluster}}$, this would lead to more errors like in Fig.~\ref{fig:set2-error-false} as a side effect. There is thus a trade-off between these two types of errors.

\subsubsection{Data set III}
The poorest Jaccard coefficient (0.78) was obtained for the event shown in figure \ref{fig:set3-merge}: one cluster was missed and merged with a different cluster due to a too small bent at the vertex. The vertex was nevertheless correctly detected due to the third track that also ends at the same vertex.

\begin{figure}[t]
  \centering
  \includegraphics[width=0.4\columnwidth]{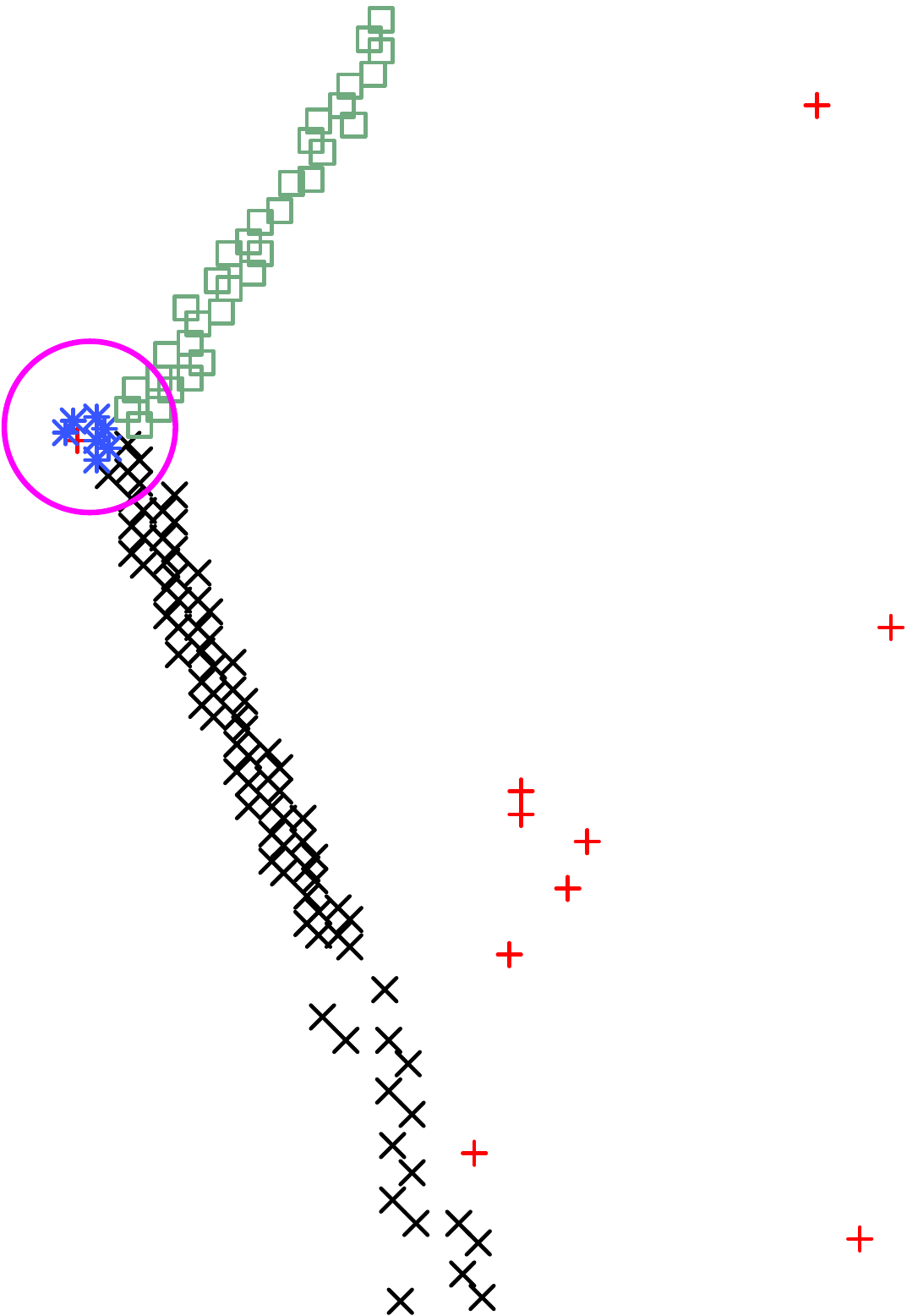}
\caption{\label{fig:set3-shortest}The shortest detected track (marked) in all events had a range of \SI{11}{mm}. Red points have been classified as noise.}
\end{figure}

From a total of 103 vertices, 99 were correctly detected, which is a recall of 96\%. In addition there were 2 false positive vertices, which leads to a precision of 98\% among the detected vertices. The shortest range among the correctly identified tracks was about \SI{11}{mm}. It is shown in figure \ref{fig:set3-shortest}.

As in this case all trajectories are straight lines, data set III provides a nice test set for comparing our new algorithm with the Hough transform, which is a standard algorithm for finding lines in point clouds \cite{dalitz17}. In order to achieve a fair comparison, we have first optimized the two parameters of the Hough transform, i.e.~the spatial cell width $dx$ and the noise threshold {\em minvotes}, such that the Jaccard coefficient was maximized. This yielded $dx=9.29$ and $\mbox{\em minvotes} = 6$. Even for this optimal parameter set, the Hough transform lead on average to a Jaccard coefficient $J=0.9045$, which is considerably smaller than the value $J=0.9706$ of our new algorithm. The difference was statistically significant for a 5\% significance level: the $p$-value of the paired $t$-test was $1.03\cdot 10^{-14}$.

\begin{figure}[t]
  \centering
  \includegraphics[width=0.6\columnwidth]{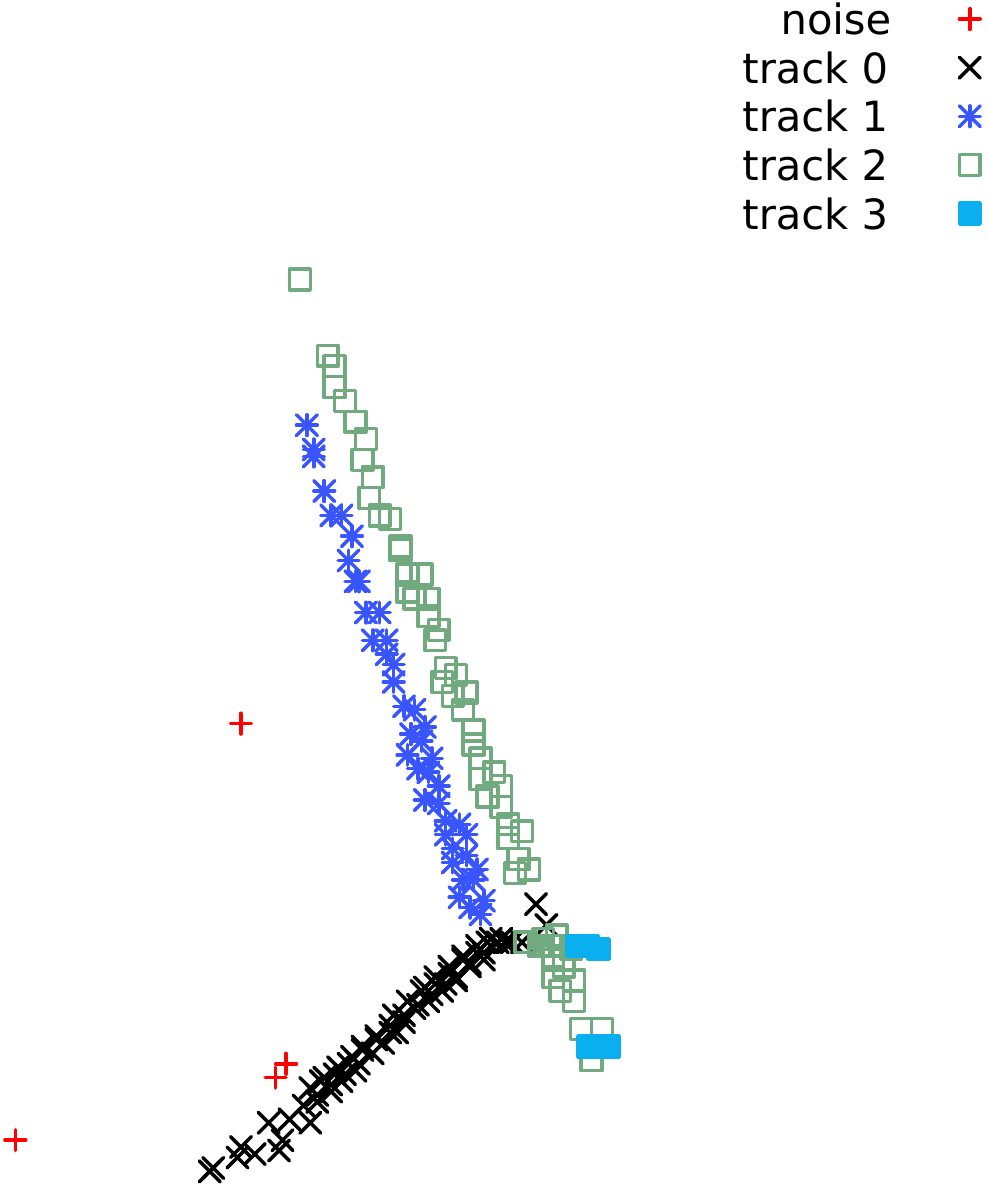}
\caption{\label{fig:houghproblem}Result of the iterative Hough transform that does not correctly identify the tracks in this event: some of the points assigned to track 0 and track 3 actually belong to track 2.}
\end{figure}

The Hough transform typically has problems with tracks that come close to each other, because the points are assigned to tracks solely based on their position. An example is shown in figure \ref{fig:houghproblem}: some points from a different track have been assigned to track 2. Our new algorithm also takes the local curve direction into account, which it is represented by the triplet orientation. This makes it possible to separate close by tracks with different orientations.

\section{Conclusions}
\label{sec:conclusions}
ATs are increasing in popularity in the low-energy nuclear physics domain as imagers for nuclear reactions due to their compelling capabilities. However, the analysis of data acquired with these detectors is a challenging task, in particular if it operates in a magnetic field. In this case, particles describe non-linear curved trajectories that form complex projection patterns, specially when the particle multiplicity is high. We have developed a novel non-parametric clustering algorithm within the context of ATs to classify and separate tracks without previous knowledge of their shape, and find the vertex of the reaction. 

The suggested algorithm for track detection was able to find most tracks and vertices on three different experimental AT data sets. In the case of absence of a magnetic field, in which case only straight tracks occur, the new algorithm performed better than an iterative Hough transform. One particular advantage of the new algorithm is that it directly returns possible vertex points as points assigned to more than one cluster. The vertex recognition rate was 86\% in the presence of a magnetic field, and 96\% in the absence of a magnetic field. Such high efficiency is mandatory for experiments where the rate of valid events is low. It is worth mentioning that the loss of efficiency caused by merging two tracks into a single one or by tracks with missing segments can be mitigated by applying simple kinematic constraints during the track reconstruction step~\cite{BRADT201765}.

A drawback of the algorithm is its dependence on appropriate choices for a several parameter values. Although we have suggested default values for these parameters, it is advisable in general to optimize the parameters on a small selection of sample events of the same reaction type. This is, however, completely compatible with the fact that for every different experiment, many physical parameters that determine the topology and structure of the hit pattern (such as drift velocity or electronics sampling rate) must be adjusted before the actual data taking process.

It should be noted that the present work defines the problem of track detection as a clustering problem, but not as a curve detection problem. The particle trajectories are left to be interpolated from each cluster, for example with the method described in \cite{lee00}. When such an interpolation is done, some clusters might represent physically improbable or even impossible trajectories. This could be a useful information that might be used as feedback to the algorithm for modifying parameters on the fly.

The promising results obtained in this work suggest that this algorithm is a powerful tool to analyze data taken with any of the ATs detectors that are operational in several nuclear physics facilities around the world~\cite{BeceiroNovo2015124}

\section*{Acknowledgments}
The AT-TPC at the NSCL was partially supported by the National Science Foundation (NSF) under grant no.~MRI-0923087. The commissioning of the AT-TPC was supported by the NSF under cooperative agreement no.~PHY-1102511.


\bibliographystyle{elsarticle-num} 
\bibliography{tpctracks-arxiv-1.2}

\end{document}